\documentstyle[12pt,epsf,rotate]{article}

\def\npb#1#2#3{{Nucl. Phys. {\bf B~#1} #2 (#3)}}
\def\plb#1#2#3{{Phys. Lett. {\bf B~#1} #2 (#3)}}
\def\prd#1#2#3{{Phys. Rev. {\bf D~#1} #2 (#3)}}
\def\prl#1#2#3{{Phys. Rev. Lett. {\bf #1} #2 (#3)}}

\def\ie{{\it i.e.}}

\def\ri{\rightarrow}
\def\ov{\overline}

\def\gsim{\ \rlap{\raise 2pt \hbox{$>$}}{\lower 2pt \hbox{$\sim$}}\ }
\def\lsim{\ \rlap{\raise 2pt \hbox{$<$}}{\lower 2pt \hbox{$\sim$}}\ }

\def\Vud{V_{ud}^{\phantom{*}}}

\def\Vuss{V_{us}^*}
\def\prk{p^{\rho}_K}
\def\psk{p^{\sigma}_K}
\def\pmk{p^{\phantom{l}}_{K\mu}}
\def\pnk{p^{\phantom{l}}_{K\nu}}
\def\pr0{p^{\rho}_0}
\def\ps0{p^{\sigma}_0}
\def\pm0{p^{\phantom{l}}_{0\mu}}
\def\pn0{p^{\phantom{l}}_{0\nu}}
\begin{document}

\sloppy

\centerline{\Large {\bf \mbox{\boldmath $K_L \rightarrow
\pi^0 \gamma e^+ e^-$} and its relation to CP and chiral tests}}
\bigskip\smallskip

\centerline{\large{John F. Donoghue and Fabrizio Gabbiani}}
\bigskip\smallskip

\centerline{\large{Department of Physics and Astronomy}}
\smallskip

\centerline{\large{University of Massachusetts, Amherst, MA ~01003}}

\vskip 3.0 truecm

\begin{abstract}

The decay $K_L \rightarrow \pi^0 \gamma e^+ e^-$ occurs at a {\it
higher} rate than the nonradiative process $K_L \rightarrow \pi^0 e^+
e^-$, and hence can be a background to CP violation studies using
the latter reaction. It also has an interest in its own right in the
context of chiral perturbation theory, through its relation to the
decay $K_L \rightarrow \pi^0 \gamma\gamma$. The leading order chiral
loop contribution to $K_L \rightarrow \pi^0 \gamma e^+ e^-$, including
the $(q_{e^+}+q_{e^-})^2/m^2_{\pi}$ dependence, is completely
calculable. We present this result and also include the higher order
modifications which are required in the analysis of $K_L \rightarrow
\pi^0 \gamma \gamma$.

\end{abstract}
{\vfill UMHEP-436}
\eject

\section{Introduction}

There are three rare decay modes of the long lived kaon which have
interrelated theoretical issues: $K_L \rightarrow \pi^0 \gamma
\gamma$, $K_L \rightarrow \pi^0 e^+e^-$ and $K_L \rightarrow \pi^0
\gamma e^+e^-$. The first two have been extensively studied; the
latter has not been previously calculated. It is the purpose of this
paper to provide a calculation of the latter process and describe how
it is related to the phenomenology of the other two decays.

There is a curious and important inverted hierarchy of these decay
modes. The rate for the radiative decay $K_L \rightarrow \pi^0
\gamma e^+e^-$ is a power of $\alpha$ {\it larger} than the
nonradiative transition $K_L \rightarrow \pi^0 e^+e^-$. This is
because the $K_L \rightarrow \pi^0 e^+e^-$ transition occurs only
through a two-photon intermediate state, or alternatively through
a one-photon exchange combined with CP violation (which numerically
appears to be roughly of the same size as the two-photon
contribution) \cite{DG}. The $K_L \rightarrow \pi^0 e^+e^-$ rate is then of
order $\alpha^4$. However, in $K_L \rightarrow \pi^0 \gamma e^+e^-$ we
need only a one-photon exchange to the $e^+e^-$, leading to a rate of
order $\alpha^3$. Our attention was first called to this inverted
hierarchy by an observation that there are infrared divergences in a
detailed study of the $K_L \rightarrow \pi^0 e^+e^-$ two-photon effect
\cite{DG} which need to be canceled by the one-loop corrections to
the radiative mode $K_L \rightarrow \pi^0 \gamma e^+e^-$ through the
contributions of the soft radiative photons. This implies that the
theoretical {\it and experimental} analyses of $K_L \rightarrow \pi^0
e^+e^-$ and $K_L \rightarrow \gamma \pi^0 e^+e^-$are tied together.
The soft and collinear photon regions of $K_L \rightarrow \gamma \pi^0
e^+e^-$ form potential backgrounds to the studies of CP violation in
the $K_L \rightarrow \pi^0 e^+e^-$ mode.

The $K_L \rightarrow \pi^0 \gamma e^+e^-$ mode also has an interest of
its own. In recent years there have been important phenomenological
studies of $K_L \rightarrow \pi^0 \gamma\gamma$  in connection with chiral
perturbation theory (ChPTh). This decay is calculable at one-loop
(\ie, order E$^4$) ChPTh with no free parameters, yielding a very
distinctive spectrum and a definite rate \cite{EPR}. Surprisingly, when the
experiment was performed the spectrum was confirmed while the measured
rate was more than a factor of 2 larger than predicted. The way out
of this problem appears to have been provided by Cohen, Ecker, and Pich
(CEP) \cite{CEP}. By adding an adjustable new effect at order E$^6$,
as well as including known corrections to the $K_L \ri \pi\pi\pi$
vertex, they found that the predicted rate can be increased
dramatically without modifying the shape of the spectrum much. This is
also a surprising result, yet as far as we know it is the unique
solution to the experimental puzzle. The ingredients of the mode
studied in this paper, $K_L \rightarrow \pi^0 \gamma e^+e^-$, are the
same as for $K_L \rightarrow \pi^0 \gamma\gamma$, except that one of
the photons is off shell. Within the framework of the CEP calculation,
the ingredients enter with different relative weights for off-shell photons.
This will allow us to test the consistency of the theoretical
resolution proposed for $K_L \rightarrow \pi^0 \gamma\gamma$.

We outline the computation for the $\cal{O}$(E$^4$) contribution to
the process in Sec. II, and then we extend it to $\cal{O}$(E$^6$)
in Sec. III. Finally, we recapitulate our conclusions in Sec. IV.

\section{The $\cal{O}$\mbox{\boldmath (E$^4$)} calculation}

First let us provide the straightforward $\cal{O}$(E$^4$) calculation
within ChPTh. This is the generalization to $k^2_1 \neq 0$ of the
original chiral calculation of EPR \cite{EPR}. Here $k_1$ is the
momentum of the off-shell photon. This captures all the
$k^2_1/m^2_{\pi}$ and $k^2_1/m^2_K$  variations of the amplitudes at this
order in the energy expansion. There can be further $k^2_1$/(1 GeV)$^2$
corrections which correspond to $\cal{O}$(E$^6$) and higher. The
easiest technique for this calculation uses the basis where the kaon
and pion fields are transformed so that the propagators have no
off-diagonal terms, as described in ref. \cite{EPR}. The relevant diagrams
are then shown in Fig. 1. Defining $\ov g$ as

\begin{equation}
{\ov g} = G^{\phantom{l}}_8/3, \qquad\qquad
G^{\phantom{l}}_8 = G^{\phantom{l}}_F \vert \Vud \Vuss \vert
g^{\phantom{l}}_8, \qquad\qquad \vert g^{\phantom{l}}_8 \vert \approx 5.1,
\end{equation}

\noindent the diagrams give the following integrals, respectively:

\begin{equation}
{\cal M}^a_{\mu\nu} = 2 e^2 {\ov g} g_{\mu \nu}
\int{{d^4 l}\over {(2\pi)^4}}
{{3 [(p^{\phantom{l}}_K-p^{\phantom{l}}_0)^2-m^2_{\pi}] - 2
[(l^2-m^2_{\pi})+(l-k_1-k_2)^2-m^2_{\pi}]} \over
{(l^2-m^2_{\pi})[(l-k_1-k_2)^2-m^2_{\pi}]}},
\end{equation}

\begin{eqnarray}
{\cal M}^b_{\mu\nu} &=& -e^2 {\ov g} \int{{d^4 l}\over {(2\pi)^4}}
{{3 [(p^{\phantom{l}}_K-p^{\phantom{l}}_0)^2-m^2_{\pi}] - 2 [(l+k_1)^2 -m^2_{\pi} + (l-k_2)^2
-m^2_{\pi}]} \over
{(l^2-m^2_{\pi})[(l+k_1)^2 - m^2_{\pi}][(l-k_2)^2 - m^2_{\pi}]}}
\nonumber \\
&\times& (2 l + k_1)_{\mu} (2 l -k_2)_{\nu}
+ (k_1, \mu) \leftrightarrow (k_2, \nu),
\end{eqnarray}

\begin{equation}
{\cal M}^c_{\mu\nu} = 8 e^2 {\ov g} g_{\mu \nu} \int{{d^4 l}\over {(2\pi)^4}}
{1 \over {l^2-m^2_{\pi}}},
\end{equation}

\begin{equation}
{\cal M}^d_{\mu\nu} = -4 e^2 {\ov g} \int{{d^4 l}\over {(2\pi)^4}} \left\{
{{(2 l -k_1)_{\mu} (2 l -k_1)_{\nu}} \over {(l^2-m^2_{\pi})[(l-k_1)^2 -
m^2_{\pi}]}} + {{(2 l -k_2)_{\mu} (2 l -k_2)_{\nu}} \over
{(l^2-m^2_{\pi}) [(l-k_2)^2 - m^2_{\pi}]}}\right\}.
\end{equation}

\begin{figure}[t]
\centering
\leavevmode
\centerline{
\epsfbox{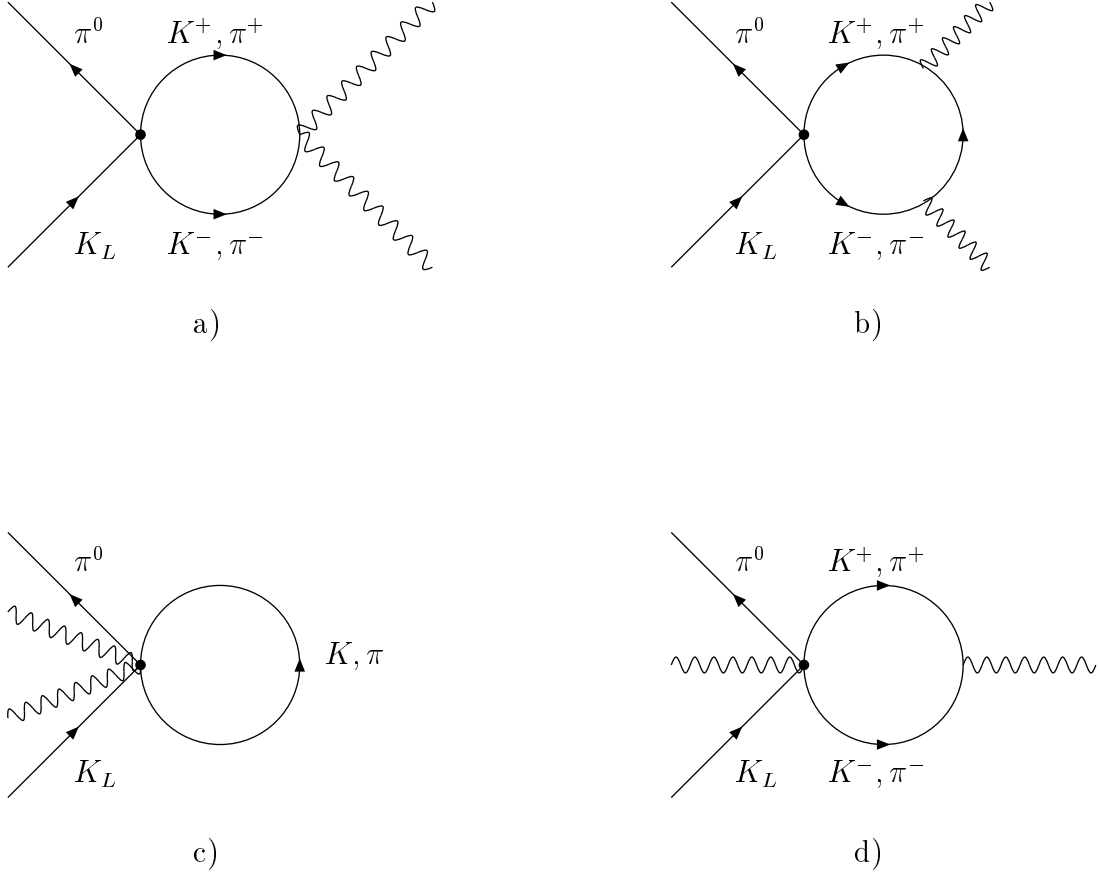}}
\caption{Diagrams relevant to the process $K_L \rightarrow
\pi^0 \gamma e^+ e^-$ at ${\cal O}(E^4)$ and ${\cal O}(E^6)$.}
\end{figure}

\noindent Interestingly when we add these together the $K \ri 3\pi$ amplitude
factors out from the remaining loop integral resulting in

\begin{equation}
{\cal M}^{\pi}_{\mu\nu} = 6 e^2 {\ov g} [(p^{\phantom{l}}_K-
p^{\phantom{l}}_0)^2-m^2_{\pi}]
\int{{d^4 l}\over {(2\pi^4)}} {{[g_{\mu\nu}(l^2-m^2_{\pi})-
(2 l +k_1)_{\mu} (2 l -k_2)_{\nu}]} \over {{(l^2-m^2_{\pi})
[(l+k_1)^2 - m^2_{\pi}] [(l-k_2)^2 - m^2_{\pi}]}}}.
\end{equation}

\noindent It is not hard to verify that this result satisfies the
constraints of gauge invariance $k^{\mu}_1 \cal{M}_{\mu\nu}$ =
$k^{\nu}_2 \cal{M}_{\mu\nu}$ = 0. At this stage, the integral may be
parametrized and integrated using standard Feynman-diagram techniques.
Let us keep photon number one as the off-shell photon and set $k^2_2 =
0$. In this case the amplitude with one photon off-shell is described
by

\begin{equation}
{\cal M}^{\pi}_{\mu\nu} = 6 e^2 {\ov g}
[(p^{\phantom{l}}_K-p^{\phantom{l}}_0)^2-m^2_{\pi}]
\left({-i} \over {16 \pi^2}\right){{(g_{\mu\nu} k_1 \cdot k_2-k_{2\mu}
k_{1\nu})} \over {k_1 \cdot k_2}} [1 + 2 I(m^2_{\pi})],
\end{equation}

\noindent with

\begin{eqnarray}
I(m^2_{\pi}) &=& \int^1_0 dz_1 \int_0^{1-z_1} dz_2
{{m^2_{\pi}-z_1(1-z_1)k^2_1} \over {2 z_1 z_2 k_1 \cdot k_2
+z_1(1-z_1)k^2_1 -m^2_{\pi} +i\epsilon}} \nonumber \\
&=& {m^2_{\pi} \over {s -k^2_1}} [F(s) - F(k^2_1)] - {k^2_1 \over {s
-k^2_1}} [G(s) - G(k^2_1)].
\end{eqnarray}

\noindent The notation is defined by

\begin{equation}
s = (p^{\phantom{l}}_K-p^{\phantom{l}}_0)^2=(k_1 + k_2)^2
\end{equation}

\noindent and

\begin{equation}
F(a) = \int^1_0 {dz_1 \over z_1} \log\left[{{m^2_{\pi} - a(1-z_1)z_1 -
i \epsilon} \over {m^2_{\pi}}}\right],
\end{equation}

\begin{equation}
G(a) = \int^1_0 dz_1 \log\left[{{m^2_{\pi} - a(1-z_1)z_1 -
i \epsilon} \over {m^2_{\pi}}}\right].
\end{equation}

\noindent The above functions are related to those presented by CEP
\cite{CEP}:

\begin{equation}
F(a) = {a \over {2 m^2_{\pi}}} \left[F_{\rm CEP}\left({a \over
4 m^2_{\pi}}\right)-1\right],
\end{equation}

\begin{equation}
G(a) = -{a \over {2 m^2_{\pi}}} \left[R_{\rm CEP}\left({a \over
4 m^2_{\pi}}\right)+{1 \over 6}\right],
\end{equation}

\noindent remembering that

\begin{eqnarray}
F_{\rm CEP}(x) & = & 1 - {1 \over x} \left[ \sin^{-1} \left(
\sqrt{x} \right) \right]^2, \, \qquad  x \leq 1, \nonumber \\
& = & 1 + {1 \over 4x} \left[ \log {1 - \sqrt{1 - 1/x} \over 1 + \sqrt{1 -
1/x}}  + i \pi \right]^2, \, \qquad x \geq 1, \nonumber \\
R_{\rm CEP}(x) & = & - {1 \over 6} + {1 \over 2x} \left[ 1 - \sqrt{1/x
- 1} \sin^{-1} \left( \sqrt{x} \right) \right], \,
\qquad x \leq 1, \nonumber \\
& & - {1 \over 6} + {1 \over 2x} \left[ 1 + \sqrt{1 - 1/x} \left( \log {1 -
\sqrt{1 - 1/x} \over 1 + \sqrt{1 - 1/x}} + i \pi \right) \right], \,
\qquad x \geq 1.
\end{eqnarray}

\noindent This agrees with the EPR result in the $k^2_1 \ri 0$ limit.

At this order we have also calculated the additional contribution
resulting from the kaons circulating in the loops of Fig. 1. They give
rise to

\begin{equation}
{\cal M}^K_{\mu\nu} = 6 e^2 {\ov g} (m^2_K+m^2_{\pi}-s)
\int{{d^4 l}\over {(2\pi^4)}} {{[g_{\mu\nu}(l^2-m^2_K)-
(2 l +k_1)_{\mu} (2 l-k_2)_{\nu}]} \over {{(l^2-m^2_K)
[(l+k_1)^2 - m^2_K] [(l-k_2)^2 - m^2_K]}}}.
\end{equation}

\noindent The resulting integral is similar to that of Eq. (8),
substituting the mass of the pion with that of the kaon. Attaching an
$e^+e^-$ couple to either photon and adding all
the above contributions together, the result we obtain for the
branching ratio is

\begin{equation}
{\rm BR}(K_L \rightarrow \pi^0 \gamma e^+e^-) = 1.0 \times 10^{-8}.
\end{equation}

\noindent With the definitions

\begin{equation}
z = {s \over {m^2_K}}, \qquad
y = {{p^{\phantom{l}}_K \cdot (k_1 - k_2)} \over {m^2_K}},
\end{equation}

\noindent the decay distributions in $z$ and $y$ provide more detailed
information. We present them in Figs. 2 and 3.

\begin{figure}[t]
\centering
\leavevmode
\epsfxsize=300pt
\epsfysize=300pt
{\centerline{\epsfbox{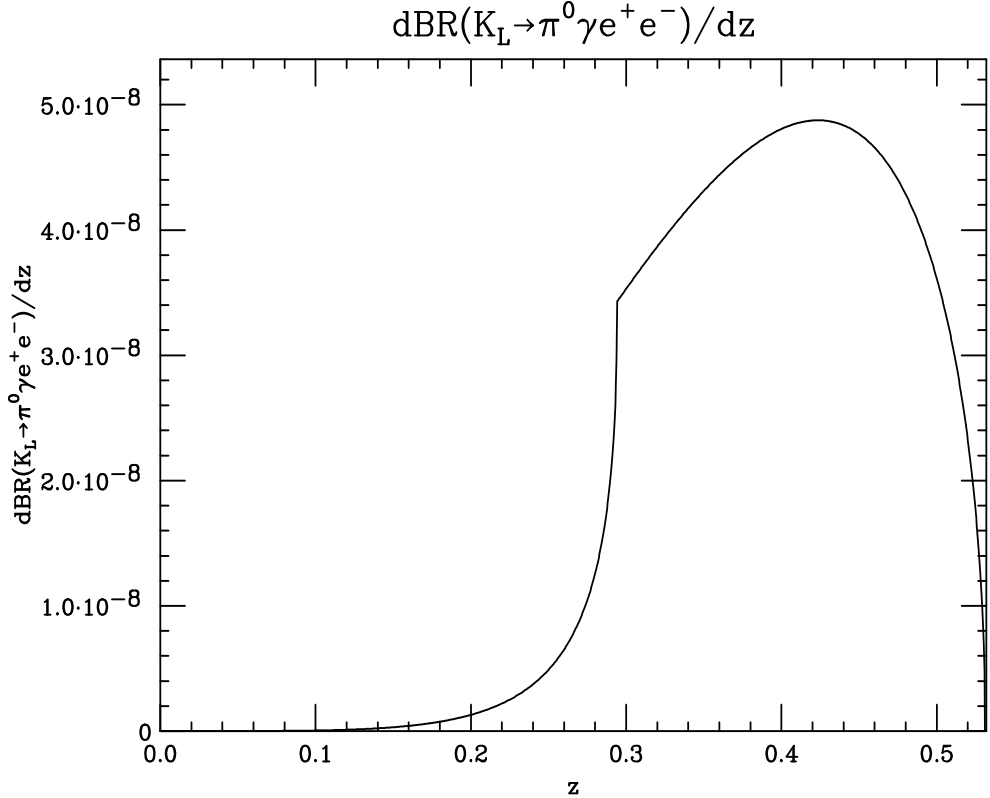}}}
\caption{Differential branching ratio $d\Gamma(K_L \rightarrow
\pi^0 \gamma e^+ e^-)/dz$ to order ${\cal O}$(E$^4$).}
\end{figure}

\begin{figure}[t]
\centering
\leavevmode
\epsfxsize=300pt
\epsfysize=300pt
{\centerline{\epsfbox{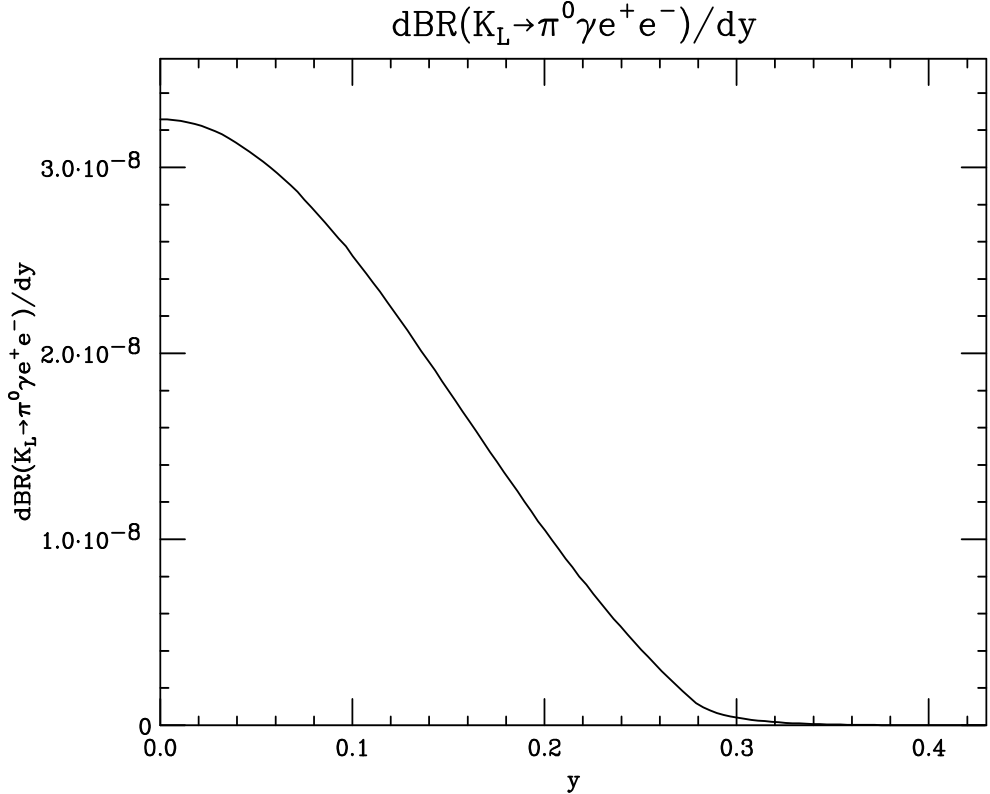}}}
\caption{Differential branching ratio $d\Gamma(K_L \rightarrow
\pi^0 \gamma e^+ e^-)/dy$ to order ${\cal O}$(E$^4$).}
\end{figure}

\section{The $\cal{O}$\mbox{\boldmath (E$^6$)} calculation}

We also wish to extend this calculation along the lines proposed by
CEP \cite{CEP}, who provide a plausible solution to the problem
raised by the experimental rate not agreeing with the $\cal{O}$(E$^4$)
calculation when both photons are on-shell. The two primary new
ingredients involve known physics which surfaces at the next order in
the energy expansion. The first involves the known quadratic energy
variation of the $K \ri 3\pi$ amplitude, which occurs from higher
order terms in the weak nonleptonic Lagrangian \cite{DGH,DA}. While the
full one-loop structure of this is known \cite{KMW}, it involves complicated
nonanalytic functions and we approximate the result at
$\cal{O}$(E$^4$) by an analytic polynomial which provides a good
description of the data throughout the physical region:

\begin{equation}
{\cal M}(K \ri \pi^+\pi^-\pi^0) = 4 a_1
p^{\phantom{l}}_K \cdot p^{\phantom{l}}_0
p^{\phantom{l}}_+ \cdot p^{\phantom{l}}_- + 4 a_2
(p^{\phantom{l}}_K \cdot p^{\phantom{l}}_+
p^{\phantom{l}}_0 \cdot p^{\phantom{l}}_- +
p^{\phantom{l}}_K \cdot p^{\phantom{l}}_-
p^{\phantom{l}}_0 \cdot p^{\phantom{l}}_+),
\end{equation}

\noindent using

\begin{equation}
a_1 = 3.1 \times 10^{-6} m^{-4}_K \qquad {\rm and}
\qquad a_2 =- 1.26 \times 10^{-6} m^{-4}_K.
\end{equation}

\noindent
$a_1$ and $a_2$ are obtained from a fit to the amplitude for $K_L \rightarrow
\pi^0\pi^+\pi^-$ \cite{DGH} and to the amplitude and spectrum for $K_L
\rightarrow
\pi^0 e^+ e^-$ \cite{CEP}, so that their values are constrained within
their theoretical uncertainty of 10 -- 20\%.  We have numerically
verified that such a variation of said parameters involves a very
modest change in the shape of the spectrum for $K_L \rightarrow \gamma \pi^0
e^+ e^-$ and a change in its final branching ratio somewhat smaller
than the uncertainty on the parameters.

\noindent The other ingredient involves vector meson exchange such as
in Fig. 4. Some of such contributions are known, but there are others
such as those depicted in Fig. 5 which have the same structure but an
unknown strength, leaving the total result unknown. In Ref. \cite{CEP}
the result is parametrized by a ``subtraction constant'' which must be
fit to the data.

\begin{figure}[t]
\centering
\leavevmode
\centerline{
\epsfbox{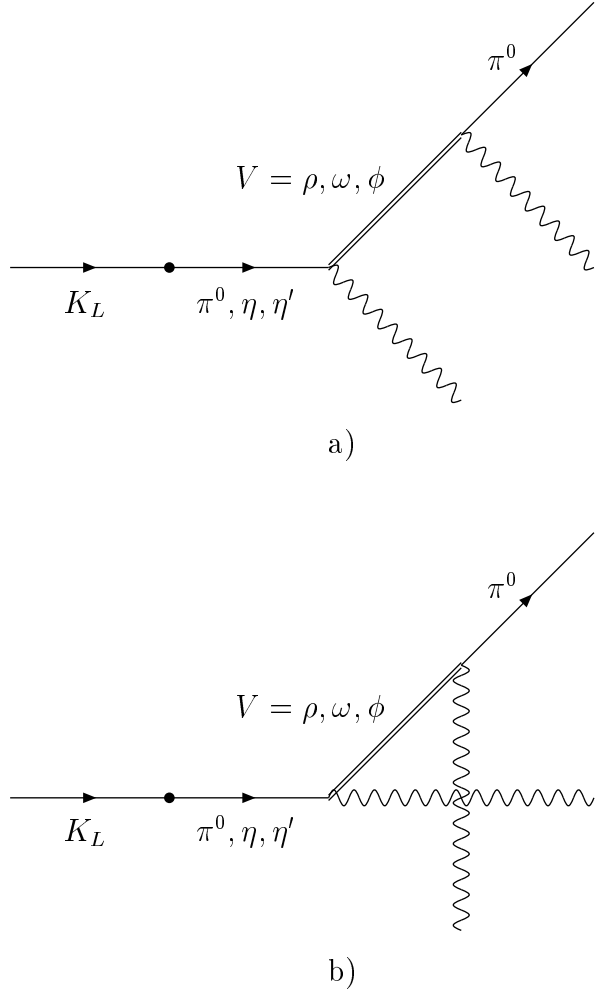}}
\caption{Vector meson exchange diagrams contributing to
$K_L \rightarrow \pi^0 \gamma e^+ e^-$.}
\end{figure}

\begin{figure}[t]
\centering
\leavevmode
\centerline{
\epsfbox{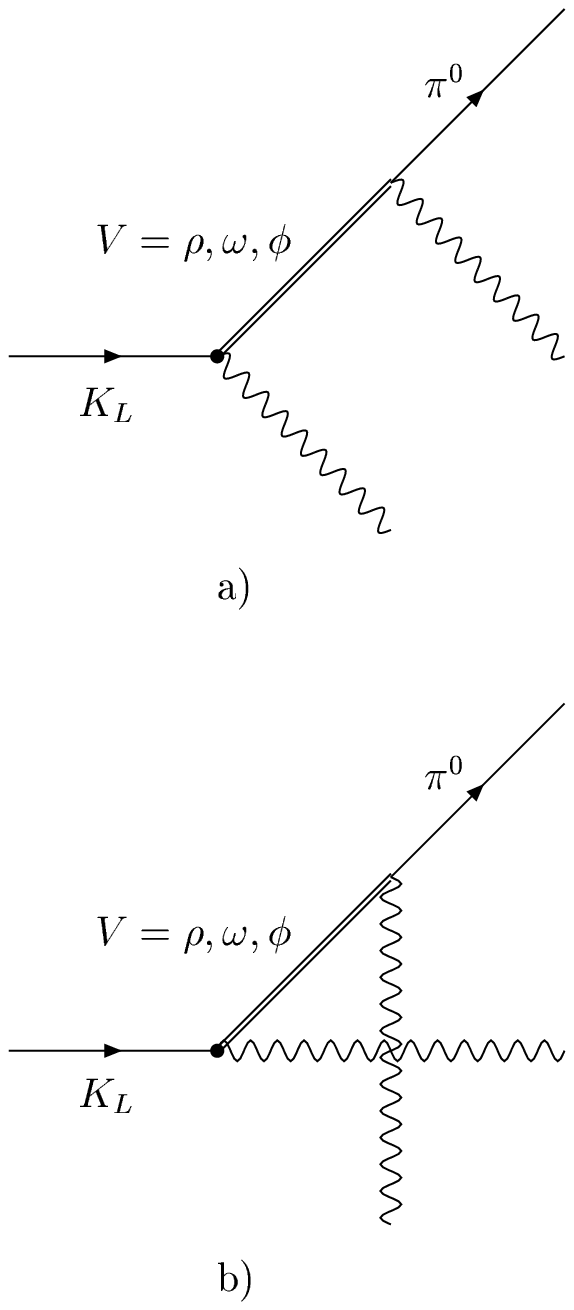}}
\caption{Vector meson exchange diagrams contributing to
$K_L \rightarrow \pi^0 \gamma e^+ e^-$ with unknown strength.}
\end{figure}

In principle one can add the ingredients to the amplitudes and perform
a dispersive calculation of the total transition matrix element. In
practice it is simple to convert the problem to an effective field
theory and and do a Feynman diagram calculation which will yield the
same result. We follow this latter course.

The Feynman diagrams are the same as shown in Fig. 1, although the
vertices are modified by the presence of $\cal{O}$(E$^4$) terms in the
energy expansion. Not only does the direct $K \ri 3\pi$ vertex change
to the form given in Eq. (18), but also the weak vertices with one and two
photons have a related change. The easiest way to determine these is
to write a gauge invariant effective Lagrangian with coefficients
adjusted to reproduce Eq. (18). We find

\begin{equation}
{\cal M_{\mu}}(K \ri \pi^+ \pi^- \pi^0 \gamma) = 4 a_1 e
(p^{\phantom{l}}_+ - p^{\phantom{l}}_-)_{\mu} + 4 a_2 e
(p^{\phantom{l}}_+ - p^{\phantom{l}}_-)_{\sigma}
(\ps0 \pmk + \psk \pm0),
\end{equation}

\begin{equation}
{\cal M}_{\mu\nu}(K \ri \pi^+ \pi^- \pi^0 \gamma \gamma) =
-8 a_1 e^2 g_{\mu\nu}
p^{\phantom{l}}_K \cdot p^{\phantom{l}}_0 + 8 a_2 e^2 (\pmk \pn0 + \pnk \pm0).
\end{equation}

\noindent The resulting calculation follows the same steps as
described above, but is more involved and is not easy to present in a
simple form. We have checked that our result is gauge invariant and
reduces to that of CEP in the limit of on-shell photons.

The contribution proportional to $a_1$ can be computed
analogously to those already calculated for the $\cal{O}$(E$^4$) case:

\begin{equation}
{\cal M}_{\mu\nu} = 4 a_1 e^2 (z-2 r^2_{\pi})(1+r^2_{\pi}-z)
{1 \over {(z-q)}}(g_{\mu\nu} k_1 \cdot k_2-k_{2\mu} k_{1\nu})
[1+2 I(m^2_{\pi})],
\end{equation}

\noindent where

\begin{equation}
r_{\pi} = {m_{\pi} \over m_K}, \qquad z = {s \over {m^2_K}},
\qquad q = {{k^2_1} \over {m^2_K}}.
\end{equation}

The $a_2$ part originates another set of integrals
which can be written as

\begin{equation}
{\cal M}^a_{\mu\nu} = -8 a_2 (\prk \ps0 + \psk \pr0) e^2
g_{\mu\nu}\int {{d^d l}
\over {(2\pi)^d}} {{l_{\rho}(l-k_1-k_2)_{\sigma}} \over
{(l^2-m^2_{\pi})[(l-k_1-k_2)^2 - m^2_{\pi}]}},
\end{equation}

\begin{eqnarray}
{\cal M}^b_{\mu\nu} &=& 4 a_2 (\prk \ps0 + \psk \pr0) e^2\int {{d^d l}
\over {(2\pi)^d}} \left\{ {{(2l+k_1)_{\mu} (2l-k_2)_{\nu}
(l+k_1)_{\rho} (l-k_2)_{\sigma}} \over
{(l^2-m^2_{\pi})[(l+k_1)^2 - m^2_{\pi}][(l-k_2)^2 -
m^2_{\pi}]}} \right. \nonumber \\
&+& \left. {{(2l+k_2)_{\nu} (2l-k_1)_{\mu}
(l+k_2)_{\rho} (l-k_1)_{\sigma}} \over
{(l^2-m^2_{\pi})[(l-k_1)^2 - m^2_{\pi}][(l+k_2)^2 -
m^2_{\pi}]}} \right\} \nonumber \\
&=& 8 a_2 (\prk \ps0 + \psk \pr0) e^2 \int {{d^d l}
\over {(2\pi)^d}} {{(2l+k_1)_{\mu} (2l-k_2)_{\nu}
(l+k_1)_{\rho} (l-k_2)_{\sigma}} \over
{(l^2-m^2_{\pi})[(l+k_1)^2 - m^2_{\pi}][(l-k_2)^2 -
m^2_{\pi}]}},
\end{eqnarray}

\begin{equation}
{\cal M}^c_{\mu\nu} = 8 a_2 (\pmk \pn0 + \pnk \pm0) e^2 \int {{d^d l}
\over {(2\pi)^d}} {1 \over {l^2-m^2_{\pi}}},
\end{equation}

\begin{eqnarray}
{\cal M}^d_{\mu\nu} &=& -4 a_2 (\ps0 \pnk + \psk \pn0) e^2 \int {{d^d l}
\over {(2\pi)^d}} {{(2l-k_1)_{\mu} (2l-k_1)_{\sigma}} \over
{(l^2-m^2_{\pi})[(l-k_1)^2 - m^2_{\pi}]}} \nonumber \\
&-& 4 a_2 (\ps0 \pmk + \psk \pm0) e^2 \int {{d^d l}
\over {(2\pi)^d}} {{(2l-k_2)_{\nu} (2l-k_2)_{\sigma}} \over
{(l^2-m^2_{\pi})[(l-k_2)^2 - m^2_{\pi}]}}.
\end{eqnarray}

\noindent From the above formulas we obtain

\begin{eqnarray}
{\cal M}_{\mu\nu} &=& {1 \over {(4 \pi)^2}}
\left[A(x_1,x_2)(k_{2\mu} k_{1\nu}-k_1 \cdot k_2 g_{\mu\nu})
\phantom{{k^2_1} \over {k_1}} \right. \nonumber \\
&+& B(x_1,x_2) \left({p^{\phantom{l}}_K \cdot k_1
p^{\phantom{l}}_K \cdot k_2 \over {k_1 \cdot k_2}}g_{\mu\nu}+
p^{\phantom{l}}_{K\mu}p_{K\nu}-{{p^{\phantom{l}}_K \cdot k_1}
\over {k_1 \cdot k_2}} k_{2\mu} p^{\phantom{l}}_{K\nu}-
{{p^{\phantom{l}}_K \cdot k_2} \over {k_1 \cdot k_2}}
k_{1\nu} p^{\phantom{l}}_{K\mu}\right) \nonumber \\
&+& \left. D(x_1,x_2) \left(k^2_1 {p^{\phantom{l}}_K \cdot k_2
\over {k_1 \cdot k_2}}g_{\mu\nu}-
{{p^{\phantom{l}}_K \cdot k_2} \over {k_1 \cdot k_2}} k_{1\mu} k_{1\nu}+
k_{1\mu}p^{\phantom{l}}_{K\nu}-{{k^2_1} \over {k_1 \cdot k_2}}k_{2\mu}
p^{\phantom{l}}_{K\nu}\right) \right],
\end{eqnarray}

\noindent where

\begin{eqnarray}
A &=& 16 a_2 e^2 \{2 [1-2(x_1 + x_2)]I_1(z_1 z_2)+x_1 I_1(z_2)+x_2
[2 I_1(z^2_2) - I_1(z_2)+ I_1(z_1)]\} \nonumber \\
&-& 32 a_2 e^2 \{[2 x^2_1 -x_1(z+q)]
[-I_2(z^3_1 z_2)+I_2(z^2_1 z_2)] \nonumber \\
&+& [2 x_1 x_2 - x_1 (z-q)/2 -x_2 (z+q)/2]
[2 I_2(z^2_1 z^2_2)+I_2(z_1 z_2)-I_2(z^2_1 z_2) \nonumber \\
&-& I_2(z_1 z^2_2)]+[2 x^2_2-x_2 (z-q)]
[I_2(z_1 z^2_2)-I_2(z_1 z^3_2)]\} \nonumber \\
&+& {4 \over 3} a_2 e^2 \log{m^2_{\pi} \over {m^2_{\rho}}} +
(4 \pi)^2 {\rm VMD}_A, \nonumber \\
\end{eqnarray}

\begin{equation}
B = -32 a_2 e^2 I_3 +16 a_2 I_4 + {4 \over 3} a_2 e^2 (z-q) \left(
-1 + \log{m^2_{\pi} \over {m^2_{\rho}}}\right) +
(4 \pi)^2 {\rm VMD}_B,
\end{equation}

\begin{eqnarray}
D = &-& {B \over 2} + 16 a_2 e^2 [2 x_2 -(z-q)/2]
[2 I_1(z_1 z_2)-I_1(z_2)] \nonumber \\
&+& 16 a_2 e^2 (2 y -q)[I_1(z_1)-I_1(1)/2]+
4 a_2 e^2 [2 x_1-(z+q)/2] I_5 \nonumber \\
&+& (4 \pi)^2 {\rm VMD}_D,
\end{eqnarray}

\noindent with

\begin{equation}
I_1(z^n_1 z^m_2) = \int^1_0 dz_1 \int^{1-z_1}_0 dz_2 z^n_1 z^m_2
\log{D_1 \over {m^2_{\pi}}},
\end{equation}

\begin{equation}
I_2(z^n_1 z^m_2) = \int^1_0 dz_1 \int^{1-z_1}_0 dz_2 {{z^n_1 z^m_2}
\over D_1},
\end{equation}

\begin{equation}
I_3 = \int^1_0 dz_1 \int^{1-z_1}_0 dz_2 D_1 \log{D_1 \over {m^2_{\pi}}},
\end{equation}

\begin{equation}
I_4 = \int^1_0 dz_1 D_2 \log{D_2 \over {m^2_{\pi}}},
\end{equation}

\begin{equation}
I_5 = \int^1_0 dz_1 (4 z^2_1-4 z_1+1) \log{D_2 \over {m^2_{\pi}}},
\end{equation}

\noindent and

\begin{eqnarray}
D_1 &=& m^2_{\pi} - 2 k_1 \cdot k_2 z_1 z_2 - k^2_1 z_1 (1-z_1), \nonumber \\
D_2 &=& m^2_{\pi} - k^2_1 z_1 (1-z_1), \nonumber \\
x_1 &=& {{p^{\phantom{l}}_K \cdot k_1} \over {m^2_K}},
\qquad x_2 = {{p^{\phantom{l}}_K \cdot k_2} \over {m^2_K}},
\end{eqnarray}

\begin{equation}
{\rm VMD}_A(x_1,x_2) = - \sum_{V= \omega,\rho} G_V
\left[{{p^{\phantom{l}}_K \cdot (p^{\phantom{l}}_K-k_2)} \over
{(p^{\phantom{l}}_K-k_2)^2-m^2_V}}+
{{p^{\phantom{l}}_K \cdot (p^{\phantom{l}}_K-k_1)} \over
{(p^{\phantom{l}}_K-k_1)^2-m^2_V}}\right],
\end{equation}

\begin{equation}
{\rm VMD}_B(x_1,x_2) = - \sum_{V= \omega,\rho} G_V
k_1 \cdot k_2 \left[{1 \over {(p^{\phantom{l}}_K-k_2)^2-m^2_V}}+
{1 \over {(p^{\phantom{l}}_K-k_1)^2-m^2_V}}\right],
\end{equation}

\begin{equation}
{\rm VMD}_D(x_1,x_2) = \sum_{V= \omega,\rho} G_V
{{k_1 \cdot k_2} \over {(p^{\phantom{l}}_K-k_1)^2-m^2_V}},
\end{equation}

\noindent assuming the numerical values \cite{HS}

\begin{equation}
G_{\rho}m^2_K = 0.68 \times 10^{-8}, \qquad
G_{\omega}m^2_K  = -0.28 \times 10^{-7}.
\end{equation}

The loop calculation that we have just described provides all of the
off-shell dependence scaled by the pion mass, and is of the form
$k^2_1/m^2_{\pi}$. There can be an additional dependence of the form
$k^2_1/\Lambda^2$, where $\Lambda$ $\approx$ 1 GeV. We cannot provide a
model independent analysis of the latter. However, experience has
shown that most of the higher order momentum dependence is well
accounted for by vector meson exchange. Therefore we include the
$k^2_1/ \Lambda^2$ dependence which is predicted by the diagrams of
Fig. 4. One can recover the parametrization in $a_V$ neglecting the
dependence on $(p^{\phantom{l}}_K-k_1)^2$ and
$(p^{\phantom{l}}_K-k_2)^2$ in formulas (38) -- (40), and performing the
replacement \cite{HS}

\begin{equation}
{{\pi G_{\rm eff} m^2_K}
\over {2 G_8 \alpha m^2_V}} \ri a_V ,
\end{equation}

\noindent where $G_{\rm eff}$ $\approx$ $G_{\rho}$+$G_{\omega}$. This
completes our treatment of the $K_L \rightarrow
\pi^0 \gamma e^+ e^-$ amplitude.

\noindent The calculation we have presented in this section
leads to the total branching ratio of

\begin{equation}
{\rm BR}(K_L \rightarrow \pi^0 \gamma e^+e^-) = 2.3 \times 10^{-8}.
\end{equation}

\noindent The decay distributions are presented in Figs. 6 and 7.

\begin{figure}[t]
\centering
\leavevmode
\epsfxsize=300pt
\epsfysize=300pt
{\centerline{\epsfbox{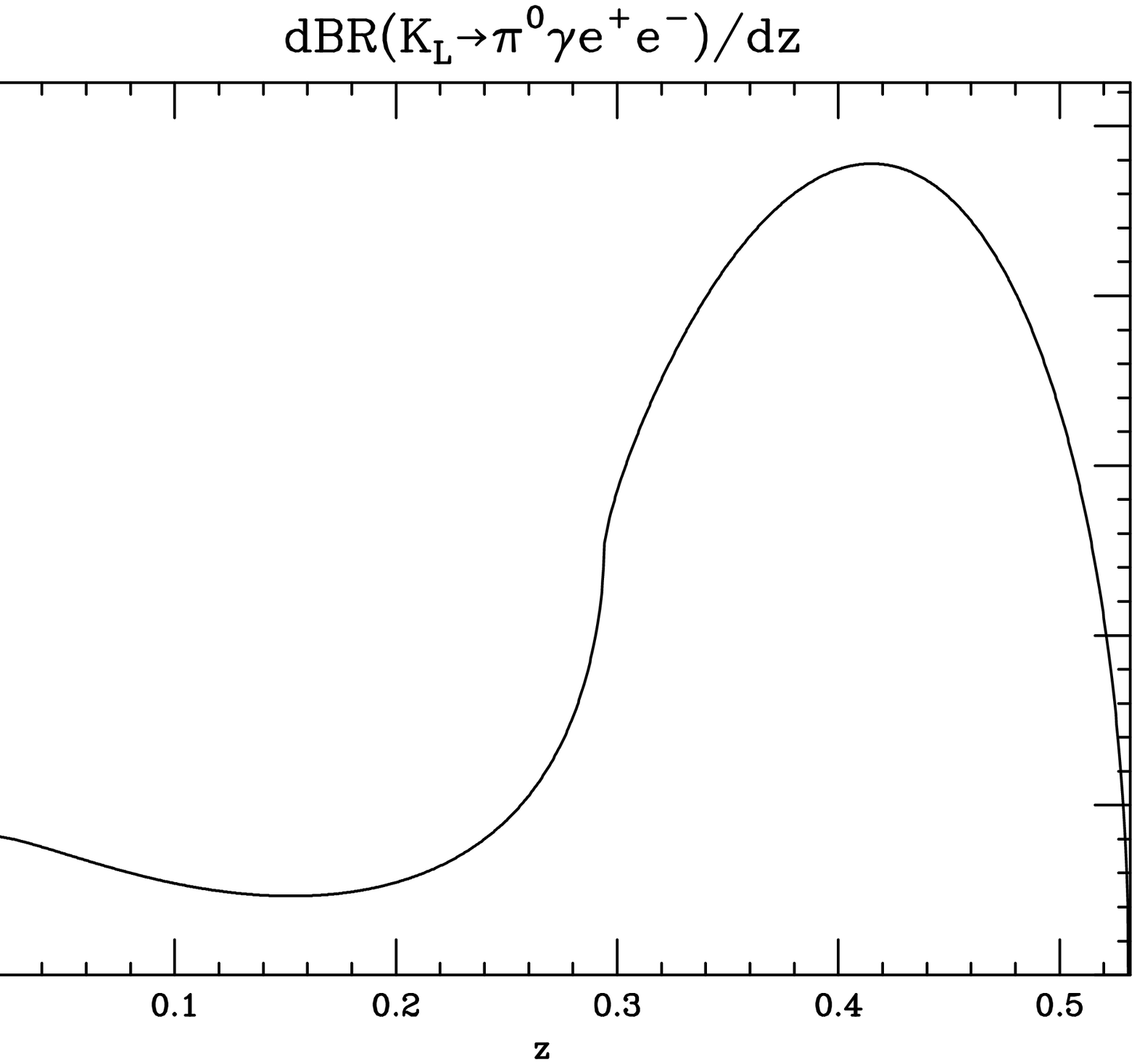}}}
\caption{Differential branching ratio $d\Gamma(K_L \rightarrow
\pi^0 \gamma e^+ e^-)/dz$ to order ${\cal O}$(E$^6$).}
\end{figure}

\begin{figure}[t]
\centering
\leavevmode
\epsfxsize=300pt
\epsfysize=300pt
{\centerline{\epsfbox{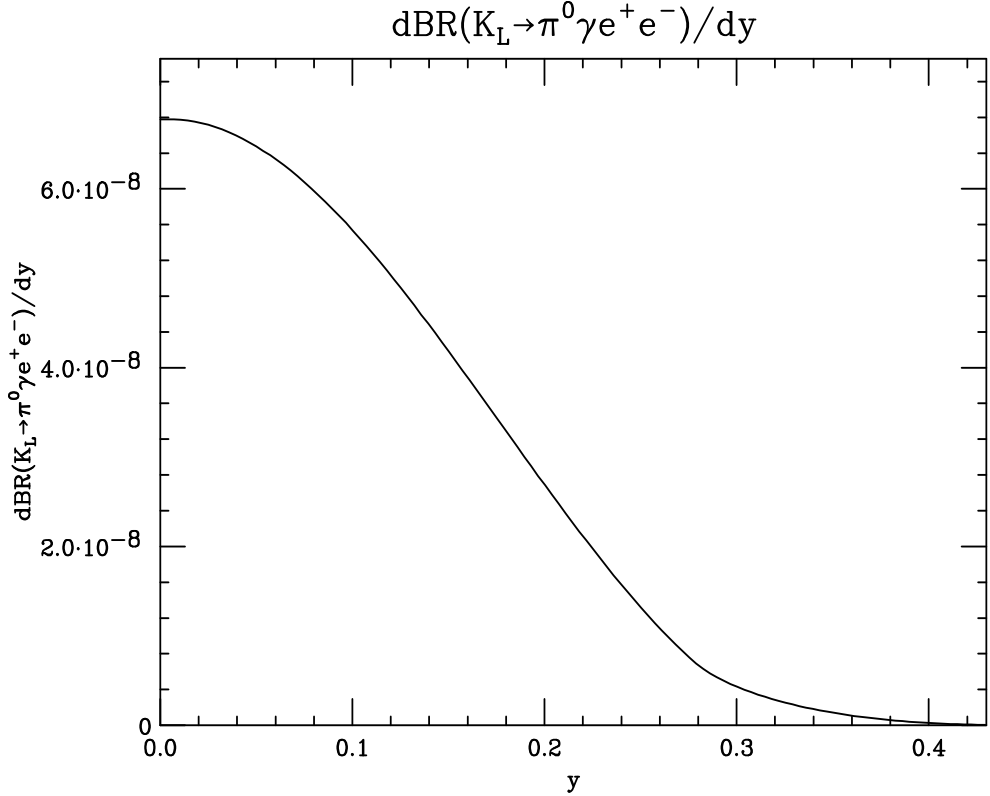}}}
\caption{Differential branching ratio $d\Gamma(K_L \rightarrow
\pi^0 \gamma e^+ e^-)/dy$ to order ${\cal O}$(E$^6$).}
\end{figure}

\section{Conclusions}

The behavior of the $K_L \rightarrow \pi^0 \gamma e^+ e^-$ amplitude
mirrors closely that of the process $K_L \rightarrow \pi^0 \gamma
\gamma$. The more complete calculation at order E$^6$ gives a rate
which is more than twice as large as the one obtained at order E$^4$,
despite the fact that the new parameter introduced at order E$^6$ is
quite reasonable in magnitude. This large change occurs partially
because the order E$^4$ calculation is purely a loop effect, while at
order E$^6$ we have tree level contributions, and loop contributions
are generally smaller than tree effects at a given order. It was more
surprising that the spectrum in $K_L \rightarrow \pi^0 \gamma
\gamma$ was not significantly modified  by the order E$^6$
contributions. These new effects are more visible in the low $z$
region of the process we have calculated, $K_L \rightarrow \pi^0
\gamma e^+ e^-$.

This reaction should be reasonably amenable to experimental
investigation in the future. It is 3--4 orders of magnitude larger
than the reaction $K_L \rightarrow \pi^0 e^+ e^-$  which is one of the
targets of experimental kaon decay programs, due to the connections of
the latter reaction to CP studies. In fact, the radiative process of
this paper will need to be studied carefully before the nonradiative
reaction can be isolated. The regions of the distributions where the
experiment misses the photon of the radiative process can potentially
be confused with $K_L \rightarrow \pi^0 e^+ e^-$ if the resolution is
not sufficiently precise. In addition, since the $\pi^0$ is detected
through its decay to two photons, there is potential confusion related
to misidentifying photons. The study of the reaction $K_L \rightarrow
\pi^0 \gamma e^+ e^-$ will be a valuable preliminary to the ultimate
CP tests.

\vfill\eject

\vfill\eject

\begin{thebibliography}{17}

\bibitem {DG}
J.~F.~Donoghue and F.~Gabbiani, \prd {51} {2187} {1995}.

\bibitem {EPR}
G.~Ecker, A.~Pich and E.~de Rafael, \npb {291} {692} {1987};
\npb {303} {665} {1988}.

\bibitem {CEP}
A.~G.~Cohen, G.~Ecker and A.~Pich, \plb {304} {347} {1993}.

\bibitem {DGH} J.~F.~Donoghue, E.~Golowich and B.~R.~Holstein, \prd {30}
{587} {1984}.

\bibitem {DA} L.~Cappiello, G.~D'Ambrosio and M.~Miragliuolo, \plb
{298} {423} {1993}; G.~D'Ambrosio, G.~Ecker, G.~Isidori and
H.~Neufeld, hep-ph/9411439, published in {\it The Second DA$\Phi$NE
Physics Handbook}, ed. L.~Maiani, G.~Pancheri and N.~Paver, INFN,
Frascati, Italy, 265 (1995); G.~D'Ambrosio and G.~Isidori,
hep-ph/9611284 (unpublished).

\bibitem {KMW} J.~Kambor, J.~Missimer and D.~Wyler, \npb {346} {17}
{1990}; J.~Kambor, J.~Missimer and D.~Wyler, \plb {261} {496} {1991};
J.~F.~Donoghue and B.~R.~Holstein, \prl {68} {1818} {1992}.

\bibitem{HS} P.~Heiliger and L.~M.~Sehgal, \prd {47} {4920} {1993}.

\end{thebibliography}
\end{document}